\documentclass[10pt,aps,prd,reprint,superscriptaddress,amsmath,amssymb,amsfonts,showkeys,noshowpacs,nofootinbib,floatfix]{revtex4-1}
\usepackage{graphicx}
\usepackage{bm}
\usepackage{calc}
\usepackage{color}
\usepackage{url}
\usepackage{array}
\usepackage{tabularx}
\usepackage{multirow}

\newif\ifAMStwofonts
\AMStwofontstrue

\def\gsim{~\rlap{$>$}{\lower 1.0ex\hbox{$\sim$}}}

\def\simpropto{\lower.2ex\hbox{$\; \buildrel \propto \over \sim \;$}}
\def\ltsim{\lower.5ex\hbox{$\; \buildrel < \over \sim \;$}}
\def\gtsim{\lower.5ex\hbox{$\; \buildrel > \over \sim \;$}}
\def\ltsim{\lower.5ex\hbox{$\; \buildrel < \over \sim \;$}}
\def\gtsim{\lower.5ex\hbox{$\; \buildrel > \over \sim \;$}}

\def\kms{\mbox{km\,s$^{-1}$}}

\def\prd{Physical Review D.}
\def\nat{Nature}
\def\astjo{Astronomical Journal}

\def\kms{\ {\rm km\,s^{-1}}}

\def\pmb#1{\setbox0=\hbox{#1}%
\kern-.025em\copy0\kern-\wd0
\kern.05em\copy0\kern-\wd0
\kern-.025em\raise.0433em\box0}

\def\vv{\boldsymbol{v}}

\def\vx{\boldsymbol{x}}

\def\vr{\boldsymbol{r}}

\def\hvr{\hat{\vr}}

\def\aj{AJ}

\def\simlt{\lower.5ex\hbox{$\; \buildrel < \over \sim \;$}}
\def\simgt{\lower.5ex\hbox{$\; \buildrel > \over \sim \;$}}

\newcommand{\beq}{\begin{equation}}
\newcommand{\eeq}{\end{equation}}
\def\beqa{\begin{eqnarray}}
\def\eeqa{\end{eqnarray}}
\def\fixit#1{}

\def\apj{ApJ}
\def\jcap{JCAP}

\def\apjs{ApJ. S}

\def\apjl{ApJL}
\def\mnras{MNRAS}
\def\aap{A\&A}

\begin{document}
\title{Speed from light: growth rate and bulk flow at \boldmath{$z\sim 0.1$}\\ from improved SDSS DR13 photometry}
\author{Martin Feix}
\email[Electronic address: ]{feix@iap.fr}
\affiliation{CNRS, UMR 7095 \& UPMC, Institut d'Astrophysique de Paris, 98 bis Boulevard Arago, 75014, Paris, France}
\author{Enzo Branchini}
\email[Electronic address: ]{branchin@fis.uniroma3.it}
\affiliation{Department of Mathematics and Physics, Universit\`a Roma Tre, Via della Vasca Navale 84, Rome 00146, Italy}
\affiliation{INFN Sezione di Roma 3, Via della Vasca Navale 84, Rome 00146, Italy}
\affiliation{INAF, Osservatorio Astronomico di Roma, Monte Porzio Catone, Italy}
\author{Adi Nusser}
\email[Electronic address: ]{adi@physics.technion.ac.il}
\affiliation{Department of Physics, Israel Institute of Technology - Technion, Haifa 32000, Israel}
\affiliation{Asher Space Science Institute, Israel Institute of Technology - Technion, Haifa 32000, Israel}
    
\begin{abstract}
Observed galaxy luminosities (derived from redshifts) hold information on the large-scale peculiar velocity field in the form of
spatially correlated scatter, which allows for bounds on bulk flows and the growth rate of matter density perturbations using large
galaxy redshift surveys. We apply this luminosity approach to galaxies from the recent SDSS Data Release 13. Our goal is twofold.
First, we take advantage of the recalibrated photometry to identify possible systematic errors relevant to our previous analysis of
earlier data. Second, we seek improved constraints on the bulk flow and the normalised growth rate $f\sigma_{8}$ at $z\sim 0.1$. Our
results confirm the robustness of our method. Bulk flow amplitudes, estimated in two redshift bins with $0.02<z_{1}<0.07<z_{2}<0.22$,
are generally smaller than in previous measurements, consistent with both the updated photometry and expectations for the $\Lambda$CDM
model. The obtained growth rate, $f\sigma_{8}=0.48\pm 0.16$, is larger than, but still compatible with, its previous estimate, and
closer to the reference value of Planck. Rather than precision, the importance of these results is due to the fact that they follow
from an independent method that relies on accurate photometry, which is a top requirement for next-generation photometric catalogues.
\end{abstract}

\keywords{Cosmology: theory, observations; large-scale structure of the universe; dark matter and dark energy; statistical methods; redshift surveys}
\maketitle

\section{Introduction}
\label{section1}
Cosmology is steadily maturing from a phase driven by high precision to a high-accuracy science. While statistical estimators and observational
strategies have been designed to minimise random errors, the focus is now shifting towards systematic uncertainties and their impact on the total
error budget. Various strategies have been adopted to tackle this problem. The most effective one is to estimate the same quantity using different
techniques applied to independent datasets. Considering the linear growth rate of matter density fluctuations, $f$, as an example, redshift-space
distortions (RSDs) have been recognised as the most promising method to estimate this important quantity \cite[e.g.,][]{Guz08, Percival2009} and
are now adopted as its standard probe in next-generation galaxy redshift surveys \citep[e.g.,][]{Levi2013, euclid2011}. Peculiar velocities measured
from distance indicators have traditionally represented an alternative probe which has provided a precious, though very local, robustness test for
analyses based on RSDs. More recently, yet another technique has been proposed to infer the growth rate from a photometric dataset complemented with
spectroscopic redshift information \citep{Nusser2012, Feix2015}. Thanks to this luminosity method, we now have independent and consistent $f$-estimates
out to redshifts $z\sim 0.1$.

The luminosity method can also be used to assess the coherence of the peculiar velocity field, most notably the bulk flow, i.e. the volume average
of the peculiar velocity field. Again, and this constitutes a second example, this technique has provided an important consistency check to other
estimates based on galaxy peculiar velocities and contributed to rule out claims of anomalously large flows that would prove difficult to justify
within the standard cosmological $\Lambda$CDM scenario \citep{Nusser2011, Branchini2012, Feix2014}.

The purpose of this paper is to further the luminosity method by reducing the impact of systematics and to improve estimates of the bulk flow and
the cosmic growth rate at $z\sim 0.1$ obtained in \cite{Feix2015}. We are able to achieve this goal using the new Sloan Digital Sky Survey Data Release
13 (SDSS DR13) catalogue \citep{York2000, SDSSDR13} in which photometry has been recalibrated to a nominal mmag level \citep{Finkbeiner2016}. Accurate
photometry is of paramount importance to bulk flows inferred from the luminosity method adopted here since small, but spatially correlated errors
can mimic spurious coherent flows and lead to biased measurements.

We aim at two main goals. Firstly, thanks to the reduction of systematic errors, we will test the robustness of our previous results. This mainly
applies to the bulk flow estimate since the measurement of the growth rate involves additional information (spatial clustering from spectroscopic
redshift surveys) which reduces the impact of systematics. Secondly, due to the fact that statistical uncertainties of the recalibrated magnitudes
are also smaller, we will be able to improve the accuracy in the bulk flow estimate.

The paper is structured as follows: after a short recap of the luminosity methodology and its underlying equations in section \ref{section2},
we introduce the SDSS data and mock galaxy samples used by our analysis in section \ref{section3}. Considering galaxies in two different
redshift bins, we present new bulk flow measurements and discuss their interpretation using mock catalogues in section \ref{section4a}. Adopting
clustering-based reconstructions of the linear velocity field, we provide updated constraints on the growth rate of density perturbations and
compare these to previous findings in section \ref{section4b}. Throughout this work, we will closely follow the notation of \cite{Feix2014,
Feix2015}, and assume a flat $\Lambda$CDM cosmology with fixed density parameters based on the Wilkinson Microwave Anisotropy Probe (WMAP) taken
from \cite{Calabrese2013}. Galaxy redshifts are expressed relative to the rest frame of the cosmic microwave background (CMB) using the dipole
estimate of \cite{Fixsen1996}.

\section{Methodology}
\label{section2}
To lowest order in linear perturbation theory, observed galaxy redshifts $z$ differ from their actual cosmological distances or redshifts $z_{c}$,
which are defined for an unperturbed background, according to \citep[e.g.,][]{SW}
\begin{equation}
\frac{z-z_{c}}{1+z} \approx \frac{V(t,\hvr r)}{c}.
\label{eq:sw}
\end{equation}
Here $\hvr$ is a unit vector along the line of sight to a given galaxy and $V$ denotes the physical radial peculiar velocity field which yields the
predominant contribution to this difference at sufficiently low redshifts. Consequently, observed magnitudes $M$, derived from galaxy redshifts,
are generally different from the true value $M^{(t)}$,
\begin{equation}
\begin{split}
M &= m - {\rm DM}(z) - K(z) + Q(z)\\
&= M^{(t)} + 5\log_{10}\dfrac{D_{L}(z_{c})}{D_{L}(z)},
\end{split}
\label{eq:magvar}
\end{equation}
where ${\rm DM}=25+5\log_{\rm 10}\lbrack D_{L}/{\rm Mpc}\rbrack$ is the distance modulus, $D_{L}$ denotes the luminosity distance, $m$ is the
apparent magnitude, and the functions $K(z)$ and $Q(z)$ account for $K$-corrections \citep[e.g.,][]{Blanton2007} and luminosity evolution with
redshift, respectively. The modulation of magnitudes $M-M^{(t)}$ is systematic across the sky and can be harnessed to obtain constraints on the
peculiar velocity field, using maximum-likelihood techniques \citep{TYS}. 

Detailed descriptions of the luminosity method and its various implementations are given in \cite{Nusser2011, Nusser2012} and \cite{Feix2014}. Here
we present a brief overview of the key elements. Considering a galaxy survey with magnitudes, spectroscopic redshifts, and angular positions
$\hvr_{i}$ on the sky, the starting point is to choose an appropriate model for the radial velocity field $V(\hvr,z)$ which is characterised by
a set of model parameters $\zeta_{k}$. To find an estimate of the $\zeta_{k}$, one maximises the probability of observing the data,
\begin{equation}
\begin{split}
P_{\rm tot} &= \prod\limits_{i}P\left (M_{i}\vert z_{i}, V_{i}(\lbrace\zeta_{k}\rbrace )\right )\\
&= \prod\limits_{i}\left (\phi(M_{i})\middle /\int_{M_{i}^{+}}^{M_{i}^{-}}\phi(M){\rm d}M\right ),
\end{split}
\label{eq:ml}
\end{equation}
where $V_{i}(\lbrace\zeta_{k}\rbrace )$ corresponds to the radial velocity field evaluated at the position of galaxy $i$, and redshift errors are
neglected \citep{Nusser2011, Nusser2012}. Here $\phi(M)$ denotes the galaxy luminosity function (LF) which is determined from the very same dataset,
and $M^{\pm}$ are the limiting magnitudes which depend on the survey's flux cuts and individual radial velocities field through the cosmological
redshift $z_{c}$ specified in Eq. \eqref{eq:sw}. The rationale of this approach is to find the set of model parameters which yield the minimal
spread in the observed magnitudes.

The luminosity method may be used to constrain the peculiar velocity field in different ways. In this study, we focus on two types of measurements.
To estimate cosmic bulk flows, we simply set the peculiar velocity model to $V(\hvr, z)=\hvr\cdot\vv_{B}$, where the components of the bulk flow
vector $\vv_{B}$ are the free model parameters determined by the likelihood procedure \citep{Nusser2011, Branchini2012}. Since the peculiar velocity
field is spatially coherent on large-scales, another possibility is to independently predict it from the observed galaxy distribution in redshift
space \citep{Peeb80, Nusser1994, Keselman2016}. The velocity field obtained in this way is a function of $\beta=f/b$, where $f$ is the growth rate
and $b$ is the linear bias between galaxies and total matter. Combined with the likelihood approach, this allows for constraining $\beta$ with
observed galaxy luminosities \citep{Nusser2012}.

An important element in the luminosity approach is to reliably estimate the galaxy LF from the given data. To this end, we adopt two different models of
the LF in our analysis. The first one was introduced in \cite{Branchini2012} and is based on a cubic spline. Details regarding the implementation of this
particular spline model are extensively discussed in \cite{Feix2014}. The second one assumes the well-known Schechter form which is characterised by the
usual parameters $M^{\star}$ and $\alpha^{\star}$ \citep{Sandage1979, schechter}. As the normalization of the LF cancels in the likelihood function, it
does not concern us here.

\section{Data}
\label{section3}
\subsection{SDSS DR13 galaxy catalogue}
\label{section3a}
The key asset of this work is the recently improved SDSS galaxy photometry of the publicly available Data Release 13 (DR13) \citep{SDSSDR13}.\footnote{
\url{http://www.sdss.org/dr13/}}
Considering only galaxies that are part of the SDSS legacy survey, the catalogue has a median redshift of $z\approx 0.1$ and provides magnitudes in five
different photometric bands which are corrected for Galactic extinction using the updated estimates of \cite{Schlafly2011}. Compared to previous data
releases, the SDSS photometry has been recalibrated using imaging data from the PanSTARRS1 survey \citep{Kaiser2010} yielding differences up to the
percent level \citep{Finkbeiner2016}. Analogue to our recent analysis based on galaxies from the SDSS Data Release 7 (DR7) \citep{abaz, Feix2014}, we use
Petrosian $r$-band magnitudes and select galaxies within $14.5 < m_{r} < 17.6$. Furthermore, we exclude galaxies with questionable spectroscopic redshifts
or photometry by requiring the corresponding {\tt zWarning} and {\tt PS1\_UNPHOT} flags to equal zero. Since it plays a negligible role in our analysis
which is insensitive to galaxy clustering, we make no attempt at accounting for fibre collisions. To minimise systematic effects resulting from uncertainties
in $K$-corrections or luminosity evolution, we express absolute magnitudes in the $^{0.1}r$-bandpass \citep{Blanton2003B}. Finally, we further constrain
the observed absolute magnitudes $M_{r}$ and redshifts of galaxies by imposing $-22.5 < M_{r} - 5\log_{10}h < -17.0$ and $0.02 < z < 0.22$. For our assumed
cosmology, this yields a working sample with approximately $4.5\times 10^{5}$ galaxies, comprising around $10^{5}$ galaxies less than the corresponding SDSS
DR7 sample considered in \cite{Feix2014}. This purer sample is labeled as {\tt luminosityA} and used for the bulk flow measurements presented in section \ref{section4a}.

Concerning the constraints on $\beta$, we also constructed a second flux-limited subsample, {\tt luminosityB}, which was obtained as in
\cite{Feix2015} by trimming the sample {\tt luminosityA} to the redshift range $0.06<z<0.12$ and selecting only galaxies within $-33^{\circ}<\eta <36^{\circ}$
and $-48^{\circ}<\lambda <51.5^{\circ}$, where $\eta$ and $\lambda$ are the SDSS survey latitude and survey longitude, respectively. This leads to a
spatially connected sample volume at $z=0.1$ for which reliable reconstructions of the peculiar velocity field can be obtained. The sample contains a
total of roughly $1.7\times 10^{5}$ galaxies which is around 15\% less compared to the size of the data sample used in \cite{Feix2015}.

Because we are dealing with galaxy samples limited to $z\approx 0.2$, we follow the lines of \cite{Feix2014} and assume that the luminosity evolution
depends linearly on redshift, i.e. 
\begin{equation}
\label{eq:qz}
Q(z)= Q_{0}\times (z-z_{0}), 
\end{equation}
where the pivotal redshift is set to $z_{0}=0.1$. Results from applying the luminosity method to these datasets are robust with respect to different
models for luminosity evolution as well as $K$-corrections of individual galaxies \citep{Feix2014}. Regarding the latter, we shall adopt the two-dimensional
polynomial model of \cite{Chilin2010} which yields $K$-corrections as a function of redshift and $g-r$ color. To compute limiting absolute magnitudes
$M^{\pm}$ at a given redshift $z$ in the $^{0.1}r$-bandpass, we resort to the mean $K$-correction specified in \cite{Feix2014}. Like in our previous
analyses based on DR7, galaxies are weighted by their angular completeness when calculating the total likelihood function $P_{\rm tot}$.

\begin{table*}
\caption{Comparison of bulk flow measurements obtained from the SDSS DR7 and DR13 galaxy samples in two redshift bins
for the different models of the LF described in the text. The quoted errors represent marginalised 68\% confidence intervals.}
\centering
\begin{tabular*}{0.95\linewidth}{@{\extracolsep{\fill}}lccccccc}
\noalign{\medskip}
\hline
\hline
\noalign{\smallskip}
 & \multicolumn{3}{c}{$0.02<z<0.07$} & \multicolumn{3}{c}{$0.07<z<0.22$} &
\tabularnewline
\noalign{\smallskip}
$\phi(M)$ & $v_{x}$ [km/s] & $v_{y}$ [km/s] & $v_{z}$ [km/s] & $v_{x}$ [km/s] & $v_{y}$ [km/s] & $v_{z}$ [km/s] &
\tabularnewline
\noalign{\smallskip}
\hline
\noalign{\smallskip}
Hybrid & $-227\pm 128$ & $-326\pm 113$ & $-239\pm 73$ & $-367\pm 92$ & $-439\pm 85$ & $-25\pm 71$ & \multirow{3}{*}{(DR7)}
\tabularnewline
Fixed & $-175\pm 126$ & $-278\pm 111$ & $-147\pm 58$ & $-340\pm 90$ & $-409\pm 81$ & $-45\pm 43$ &
\tabularnewline
Schechter & $-151\pm 130$ & $-277\pm 116$ & $-102\pm 78$ & $-422\pm 93$ & $-492\pm 86$ & $-150\pm 74$ &
\tabularnewline
\noalign{\medskip}
Hybrid & $-200\pm 140$ & $-292\pm 122$ & $-146\pm 84$ & $-349\pm 100$ & $-301\pm 92$ & $129\pm 87$ & \multirow{3}{*}{(DR13)}
\tabularnewline
Fixed & $-199\pm 140$ & $-292\pm 121$ & $-129\pm 61$ & $-363\pm 100$ & $-323\pm 90$ & $70\pm 47$ &
\tabularnewline
Schechter & $-202\pm 141$ & $-287\pm 124$ & $69\pm 92$ & $-356\pm 100$ & $-324\pm 92$ & $19\pm 93$ &
\tabularnewline
\noalign{\smallskip}
\hline
\hline
\end{tabular*}
\label{table1}
\end{table*}

\subsection{Mock galaxy catalogues}
\label{section3b}

To aid the interpretation of bulk flow measurements, we consider a suite of 269 mock galaxy catalogues which were modelled after the SDSS DR7 and
allow for an assessment of effects due to known systematics, incompleteness, and cosmic variance. A detailed description of how these mocks were
constructed can be found in \cite{Feix2014}. Since the number of galaxies in the DR7 and DR13 samples is comparable and the photometric recalibration
of $r$-band magnitudes in DR13 introduced only slight changes at the level of a few mmag, the mock catalogues remain a suitable choice for the present
study. Crossmatching DR13 and DR7 galaxies within one arcsec, for example, we find that more than 95\% of all galaxies in {\tt luminosityA} are included
in the corresponding DR7 sample of \cite{Feix2014}. As we show in section \ref{section4a}, a further indication is provided by a very similar LF estimate
between the two datasets.

The mock catalogues include a systematic error in the SDSS DR7 photometric calibration which results in an overall zero-point photometric tilt at the
level of 10 mmag over the entire survey area \citep{Pad2008}. This tilt was modeled by a randomly oriented dipole which is characterised by a root mean
square of $\delta m_{\rm dipole}=0.01$ using all galaxies in the sample. Since the impact of such systematics is expected to be significantly reduced
in the new SDSS DR13 calibration \citep{Finkbeiner2016}, we have removed the dipole contribution from all mocks before considering them in our analysis.

\begin{figure} 
\includegraphics[width=0.95\linewidth]{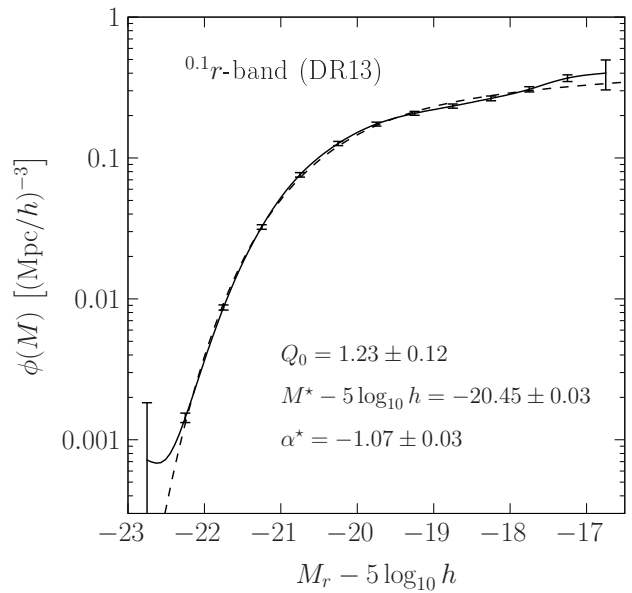}
\caption{The $^{0.1}r$-band luminosity function of the {\tt luminosityA} sample from the SDSS DR13: illustrated are the
maximum-likelihood result adopting the spline-based estimator with $\Delta M=0.5$ (solid line), and a fit based on the
Schechter model (dashed line). Error bars correspond to marginalised 99.7\% confidence limits of individual spline points.}
\label{fig1}
\end{figure}

\section{Data analysis}
\label{section4}
In what follows, we apply the luminosity method to the SDSS DR13 galaxy samples. The new bulk flow measurements are presented and
interpreted in section \ref{section4a} and updated constraints on the cosmic growth rate are discussed in \ref{section4b}. For both
cases, we will compare our results to previous estimates based on the corresponding SDSS DR7 datasets \citep{Feix2014, Feix2015}.

\subsection{Bulk flow estimates}
\label{section4a}
To begin with, we determine the LF, denoted by $\phi(M)$, in the $^{0.1}r$-band for the sample {\tt luminosityA}, assuming a vanishing
velocity field. Choosing the spline-based estimator with a spline-point separation of $\Delta M = 0.5$, the result is illustrated as the
solid line in Figure \ref{fig1}. Just as in \cite{Feix2014}, $\phi(M)$ is normalised to unity over the sample's absolute magnitude range,
and the error bars were obtained from the ``constrained'' covariance matrix which enforces the normalization constraint by a Lagrangian
multiplier \citep{James2006}. Despite a smaller estimate of the evolution parameter, $Q_{0} = 1.23\pm 0.12$ ($1.60\pm 0.11$ for DR7), the
found result agrees well with studies based on previous data releases \citep{Blanton2003, Montero2009, Feix2014}. Fitting our estimate of
$\phi(M)$ with a Schechter form, we obtain the parameters $M^{\star}-5\log_{10}h=-20.45\pm 0.03$ and $\alpha^{\star}=-1.07\pm 0.03$, which
is fully consistent with the analysis of SDSS DR7 data in \cite{Feix2014} and supports the argument in section \ref{section3b}. The
corresponding Schechter fit is shown as the dashed line in Figure \ref{fig1}. 

\begin{figure} 
\includegraphics[width=0.95\linewidth]{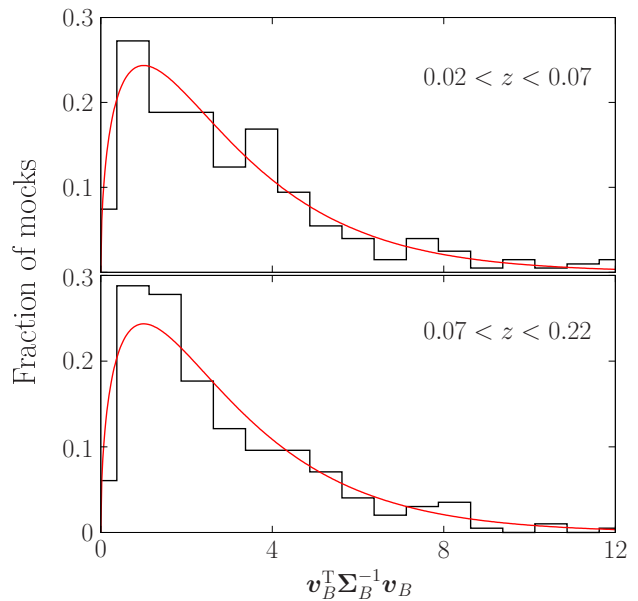}
\caption{Distribution of the quadratic form $\vv_{B}^{\rm T}\bm{\Sigma}_{B}^{-1}\vv_{B}$ in two redshift bins derived from
the SDSS mock catalogues, assuming the Schechter model of the LF: the histograms represent the observed (normalised) distribution
and solid lines correspond to the $\chi^{2}$-distribution with $k=3$ degrees of freedom.}
\label{fig2}
\end{figure}

Since the SDSS samples cover only part of the sky, bulk flow measurements generally probe linear combinations of different velocity moments
due to statistical mixing. This also includes the LF model and its evolution which may effectively introduce an additional monopole term. If interpreted
with appropriate mock catalogues mimicking the angular footprint of the real survey, however, such measurements can yield meaningful results. Here we shall
follow the strategy described in \cite{Feix2014} to estimate bulk flows in two redshift bins with $0.02 < z_{1} < 0.07 < z_{2} < 0.22$.\footnote{As
detailed in \cite{Feix2014}, this choice mainly follows from requiring comparable signal-to-noise ratios between the two redshift bins.} In particular,
we consider the following different approaches regarding the treatment of the LF in the likelihood analysis:
\begin{enumerate}
\item Estimate the LF with the spline-based model for a vanishing velocity field, and keep it fixed in the subsequent bulk flow analysis.
\item Fit a Schechter form to the spline-based LF estimate for a vanishing velocity field and model the LF as a superposition of a Schechter
form and the corresponding residual (hybrid approach).
\item Adopt a Schechter model for the LF.
\end{enumerate}
The inferred bulk flows are summarised in Table \ref{table1}. For comparison, we also list the recent estimates based on SDSS DR7 \citep{Feix2014}. The
components of the bulk flow are expressed in a Cartesian coordinate system defined by its $x$-, $y$-, and $z$-axes pointing towards Galactic coordinates
$(l,b)\approx (81^{\circ},-7^{\circ})$, $(172^{\circ},-1^{\circ})$, and $(90^{\circ},83^{\circ})$, respectively, where the $z$-axis roughly aligns
with the direction towards the centre of the northern survey region. Measurement errors were derived from the covariance matrix $\bm{\Sigma}$ which is
obtained by direct inversion of the observed Fisher matrix, defined as $\mathbf{F}_{\alpha\beta}=-\partial\log P_{\rm tot}/(\partial x_{\alpha}\partial x_{\beta})$
evaluated at the most likely parameter vector $\hat{\vx}^{\rm ML}$.

Considering total bulk flow amplitudes, the DR13 estimates are generally smaller than the DR7 ones. Averaging the results over the different
LF models yields values of $\vert\vv_{B}\vert\approx 370\pm 115$ and $485\pm 95$ in units of km~s$^{-1}$ for the low and high-redshift bin,
respectively. In the high-$z$ bin, the flow amplitudes are reduced by about 50--200~km~s$^{-1}$ compared to DR7, where the most significant
changes appear for the Schechter model of the LF. The bulk flow components found from estimators using different LF models are in reasonable
agreement, typically consistent within the quoted uncertainties. Assuming the results based on the hybrid model, the estimated bulk flows from DR13
are pointing towards $(l,b)\approx (315^{\circ},-17^{\circ})\pm (52^{\circ},15^{\circ})$ and $(304^{\circ},22^{\circ})\pm (14^{\circ},11^{\circ})$
for the first and second redshift bin, respectively. Similar directions are obtained for the other estimators.

\begin{table}
\caption{Comparison of estimated event levels (expressed in standard deviations $\sigma$) for the bulk flow amplitudes
obtained from the SDSS DR7 and DR13 galaxy samples in two redshift bins assuming different models of the LF.}
\centering
\begin{tabular*}{0.95\linewidth}{@{\extracolsep{\fill}}lccc}
\noalign{\medskip}
\hline
\hline
\noalign{\smallskip}
 & \multicolumn{2}{c}{Event level $\left\lbrack\sigma\right\rbrack$} &
\tabularnewline
\noalign{\smallskip}
$\phi(M)$ & $0.02<z<0.07$ & $0.07<z<0.22$ &
\tabularnewline
\noalign{\smallskip}
\hline
\noalign{\smallskip}
Hybrid & 1.96 & 2.75 & \multirow{3}{*}{(DR7)}
\tabularnewline
Fixed & 2.40 & 2.62 &
\tabularnewline
Schechter & 0.95 & 3.19 &
\tabularnewline
\noalign{\medskip}
Hybrid & 1.30 & 2.48 & \multirow{3}{*}{(DR13)}
\tabularnewline
Fixed & 2.26 & 2.42 &
\tabularnewline
Schechter & 1.18 & 2.22 &
\tabularnewline
\noalign{\smallskip}
\hline
\hline
\end{tabular*}
\label{table2}
\end{table}

To better assess these measurements, we repeated the analysis for the 269 mock galaxy catalogues introduced in section \ref{section3b}. In contrast
to the study presented in \cite{Feix2014}, these mocks do not account for the photometric tilt of the SDSS DR7 calibration which severely contaminates
bulk flows measured through the luminosity approach. The results based on our mocks suggest that the individual flow components obtained with our
methods are not statistically independent, but subject to correlations at the level of 0.1--0.3. If the joint distribution of bulk flow components
is approximately given by a multivariate Gaussian with covariance matrix $\bm{\Sigma}_{B}$, the quadratic form $\vv_{B}^{\rm T}\bm{\Sigma}_{B}^{-1}\vv_{B}$
should follow a $\chi^{2}$-distribution with $k=3$ degrees of freedom. To test the validity of this assumption, we computed $\vv_{B}^{\rm T}\bm{\Sigma}_{B}^{-1}\vv_{B}$
directly from the SDSS mocks. For the Schechter LF model, the resulting distributions, plotted as histograms in Figure \ref{fig2}, are indeed well-matched
by the $\chi^{2}$-distribution (solid curves). Adopting the other models of the LF yields very similar results. Using the estimate of $\bm{\Sigma}_{B}$
from the mock catalogues, we may, therefore, assign probabilities to the bulk flow measurements listed in Table \ref{table1}. As customary, we express these
in terms of confidence limits based on normally distributed data.

The corresponding event levels (in units of the standard deviation $\sigma$) for the inferred bulk flow amplitudes are presented in Table \ref{table2}.
Compared to the DR7 results, the probability estimates of the DR13 measurements are on average considerably larger, and thus consistent with an improved
photometric calibration. The most prominent change is found in the high-$z$ bin for the Schechter model of the LF, where the event level drops from
3.19 to 2.22$\sigma$. All measured DR13 flow amplitudes are contained within the $2.5\sigma$ confidence interval. Given the remaining uncertainties in
the photometric data, limitations in the modelling of the mock catalogues, and their relatively small number, we conclude that the estimated bulk flows
for SDSS DR13 are in agreement with the standard $\Lambda$CDM model. As was already pointed out in \cite{Feix2014}, we emphasize that our results are
robust with respect to the particular choices of the background cosmology\footnote{For example, varying the total matter density parameter $\Omega_{m}$
in spatially flat cosmologies over the range $0.25<\Omega_{m}<0.35$ yields almost the same estimates of the various flow components, typically within a
few $\kms$.}, $K$-corrections, and the luminosity evolution.

\subsection{Constraints on the growth rate at \boldsymbol{$z\sim 0.1$}}
\label{section4b}
Following the procedure outlined in \cite{Feix2015}, we now derive constraints on $\beta$ from the recalibrated SDSS DR13 data. The deviations in
photometry between the bulk of DR13 and DR7 galaxies are around few mmag and have only little impact on the selection of volume-limited subsamples
used to build models of the linear peculiar velocity field. Hence, we do not expect any relevant changes in the reconstruction of velocities which
is generally robust to the adopted smoothing length and features on small scales \citep{Nusser1994, Nusser2012}. For this reason, we resort to the
velocity models of \cite{Feix2015} which were derived from the SDSS DR7 catalogue for discrete and equidistant $\beta$-values with $\Delta\beta=0.05$
and $0\leq\beta\leq 1$. These models are based on a decomposition of the smoothed galaxy density field into spherical harmonics ($l_{\rm max}=150$)
and assume a Gaussian filter of $10h^{-1}$ Mpc radius. As argued in \cite{Feix2015}, the contributions due to the monopole and dipole terms are
uncertain and were removed from the velocity reconstruction ($l>1$). Using the linear velocity models, galaxies in the subsample {\tt luminosityB} 
are supplied with radial velocities $V_{i}(\beta )$ and then used in the likelihood procedure to determine the most probable $\beta$-value. To maximise
$P_{\rm tot}$, we adopt the spline-based LF which is fixed to its estimate for zero galaxy velocities. Our analysis also accounts for a slight
statistical bias which emerges from the partial sky coverage of the considered SDSS datasets. All details regarding the calculation are summarised
in \cite{Feix2014}.

\begin{figure} 
\includegraphics[width=0.95\linewidth]{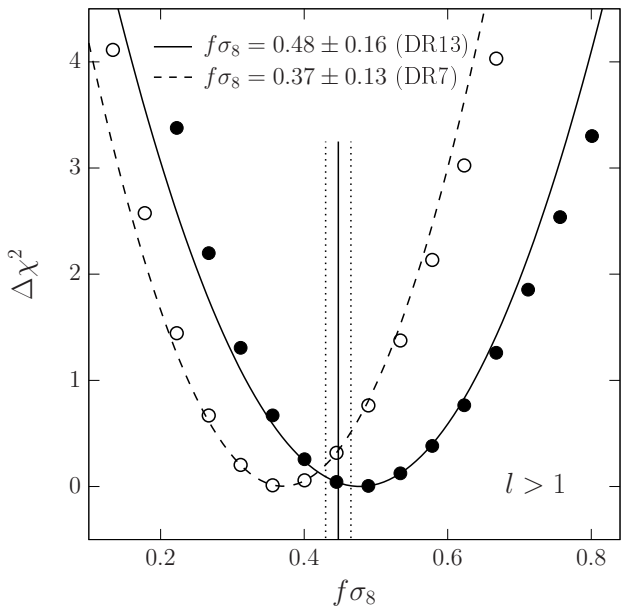}
\caption{Estimated $\Delta\chi^{2}$ (and its quadratic approximation) as a function of $f\sigma_{8}$ for linear velocity models
with $l>1$. Presented are results based on galaxies from SDSS DR13 (filled circles) and DR7 (open circles). The estimates assume
the power spectrum amplitude of $L^{\star}$-galaxies given by \cite{Tegmark2004}. Vertical lines indicate the result inferred
from Planck data (TT+lowP+lensing) \cite{Planck2016} and its 95\% confidence limits.}
\label{fig3}
\end{figure}

Carrying out the above steps, we find $\beta=0.54\pm 0.18$ for the DR13 galaxy sample. This new estimate is around 30\% larger than our
previous result ($\beta=0.42\pm 0.14$ for DR7), but consistent within the quoted uncertainties which have been obtained from approximating
the log-likelihood near its maximum to quadratic order. For velocity models which exclude more of the low multipoles (e.g., models with
$l>5$), we observe a similar trend. Assuming the power spectrum amplitude of $L^{\star}$-galaxies inferred in \cite{Tegmark2004}, we may
rewrite the result in terms of $f\sigma_{8}$, which allows for a comparison to measurements based on different datasets \citep{Song2009}
and where $\sigma_{8}$ is the amplitude of matter fluctuations in spheres of $8h^{-1}$ Mpc radius. This yields a value of $f\sigma_{8}=0.48\pm 0.16$
(and $f\sigma_{8}=0.37\pm 0.13$ for DR7; see Figure \ref{fig3}) close to the $\Lambda$CDM estimate at $z=0.1$ based on the Planck data
\citep[$f\sigma_{8}\approx 0.45$; shown as the solid vertical line in Figure \ref{fig3};][]{Planck2016}.

Our result is independent of the Hubble constant and quite insensitive to the precise choice of the cosmological parameters, the removal
of high-$l$ modes in the velocity models \citep{Feix2015}, or the used LF model. Assuming a Schechter form changes the result marginally,
i.e. $\beta=0.55\pm 0.18$. Similar to the case of bulk flows, differences in the treatment of $K$-corrections and luminosity evolution
have only a minor impact on the analysis \citep{Nusser2012, Feix2014}.

To summarise, the constraints on the large-scale peculiar velocity field at $z\sim 0.1$, derived from SDSS DR13 galaxies using the luminosity
fluctuation method, are in excellent agreement with the standard $\Lambda$CDM model of cosmology. Compared to previous results obtained
for DR7, measured bulk flow amplitudes are generally reduced, especially at higher redshifts ($0.07<z<0.22$) where photometric uncertainties
have a more significant impact on the analysis. The estimated growth rate is slightly larger than, but still compatible with, the DR7 result,
yielding a value close to the extrapolation based on Planck data. The found changes are consistent with the photometric recalibration of
$r$-band magnitudes to a nominal mmag level \citep{Finkbeiner2016} and further confirm the robustness of the luminosity method for datasets
with accurate photometry. From this point of view, all galaxy catalogues with similarly small photometric errors \citep{Abbott2016, LSST2012,
euclid2011} could be considered for future applications.

\begin{acknowledgments}
This research was supported by the I-CORE Program of the Planning and Budgeting Committee, THE ISRAEL SCIENCE FOUNDATION (grants No. 1829/12
and No. 203/09), the German-Israeli Foundation for Research and Development, the Asher Space Research Institute, and in part by the Lady
Davis Foundation. M.F. acknowledges support by the grant ANR-13-BS05-0005 of the French National Research Agency. E.B. is supported
by INFN-PD51 INDARK, MIUR PRIN 2011 ``The dark Universe and the cosmic evolution of baryons: from current surveys to Euclid'', and the Agenzia
Spaziale Italiana from the agreement ASI/INAF/I/023/12/0.
Funding for the SDSS and SDSS-II has been provided by the Alfred P. Sloan Foundation, the Participating Institutions, the National Science
Foundation, the U.S. Department of Energy, the National Aeronautics and Space Administration, the Japanese Monbukagakusho, the Max Planck
Society, and the Higher Education Funding Council for England.
\end{acknowledgments}

\bibliography{velDR13.bib}
\end{document}